\documentclass{article}
\usepackage{epsfig}
\usepackage{amssymb}
\newcommand{\bfr}{\begin{flushright}}
\newcommand{\efr}{\end{flushright}}
 
\begin{document}
\title{More on quantum kinks in gauge theories on $R^2 \otimes S^1$
}
\author{Takuya Maki\\
Department of Physics, Tokyo Metropolitan University,\\
Minami-ohsawa, Hachioji-shi, Tokyo 192-03, Japan\\
and\\
Kiyoshi Shiraishi\\
Akita Junior College, Shimokitade-Sakura, Akita-shi, Akita 010, Japan
}
\date{Il Nuovo Cimento {\bf A107} (1994) pp. 1219--1227
}
\maketitle
\begin{abstract}
Quantum kinks in gauge theories on $R^2\otimes S^1$ space-time are
studied. To obtain the, explicit profile of the kinks, we calculated
effective actions including derivative corrections for vacuum gauge
fields on $S^1$ at zero, and finite temperature up to one-loop level. It
is found that there are soliton solutions in scalar QED. We also find
that the derivative correction tends to make the vacuum unstable in
$SU(2)$ Yang-Mills theory.
\end{abstract}

\section{Introduction}
Gauge theories on space-times with non-trivial topologies have been
investigated in the literature on Kahza-Klein theories and the
low-energy superstring theories. The Wilson-line symmetry breaking is
known as an alternative mechanism to break the gauge symmetry in such
theories \cite{1}. The vacuum expectation value for the gauge field on a
compact space plays a role of a Higgs condensate in the mechanism.

Solitons in the space-time with non-trivial topology have recently been studied in
connection with topological aspects of gauge theories%
\footnote{Classical `solitons' which interpolate different gauge
vacua are studied in ref. \cite{2}.}. For QED on $R^2\otimes S^1$,
Higuchi and Parker \cite{3} have shown the existence of kink solutions,
using an effective potential for vacuum gauge field at one-loop level.

Since the kinks are the spatially varying configuration of the vacuum gauge field,
the derivative corrections to the action, which come from the loop effect, may become
important. If one wish to study the properties of the quantum solitons, one should
evaluate the derivative corrections to the effective action.

In this paper, we calculate the effective action for the vacuum gauge field,
including the second-derivative corrections for scalar QED and $SU(2)$
Yang-Mills theory on $R^2\otimes S^1$ space-time and investigate the
quantum soliton at one-loop level.

In sect. 2, we provide the effective action for the vacuum gauge field, which is
induced by complex scalar fields and Dirac fermions on $R^2\otimes S^1$.
In sect. 3, we show the explicit shape and the mass of the quantum kink
in scalar QED on $R^2\otimes S^1$. The high-temperature case is examined
in sect. 4. In sect. 5, we derive the second-derivative terms for the
vacuum gauge field in $SU(2)$ Yang-Mills theory on $R^2\otimes S^1$.
Section 6 is devoted to conclusion. 

We reserve the explicit form of the
one-loop corrections ($V$ and $\gamma$ in the text) to the appendix.

\section{Effective action}
The one-loop effective action is formally written as
\begin{equation}
\Gamma^{(1)}=\frac{1}{2}{\rm tr } \ln \Delta\,.
\label{2.1} 
\end{equation}

For a complex scalar coupled with a $U(1)$ gauge field on $R^2\otimes
S^1$, the second-order differential operator $\Delta$ is taken as
\begin{equation}
\Delta=-\partial^2_t-\partial^2_x-(\partial_y\pm ie\langle A_y
\rangle)^2\,,
\label{2.2}
\end{equation}
where the coordinate $y$ corresponds to the compactified dimension and
$e$ is the gauge coupling. We will work with the Euclidean signature for
the metric. We will also introduce the dimensionless variable $\phi=
eL\langle A_y\rangle$. We treat $\phi$ as a spatially ($x$-) dependent
quantity.

To calculate the effective action for spatially dependent background, we adopt the
method introduced by Moss, Toms and Wright \cite{4}. The Green's
function at the zeroth order in the derivative expansion \cite{4} in
the momentum space is given by 
\begin{equation}
G_0=(k^2+(2\pi l\pm\phi)^2/L^2)^{-1}\,,
\label{2.3} 
\end{equation}
where $l$ is an integer, $L$ is the length of the circumference of
$S^1$. Since further procedure to obtain the derivative corrections to
the effective action is similar to the way used by Moss, Toms and Wright
\cite{4}, we will not repeat it here. We note however that there is a
large difference in the manner of taking the trace with respect to the
momentum space, because of the intertwining between vacuum gauge field
$\phi$ and the discrete momentum on
$S^1$.

We consider the $U(1)$ gauge theory coupled to $N$ `copies', of complex
scalar fields. The Lagrangian for this case is
\begin{equation}
L_{\mbox{scalar QED}}=\frac{1}{4}F^{\mu\nu}F_{\mu\nu}+\sum_{i=1}^N
\left|\left(\partial_\mu-i\frac{e}{\sqrt{N}}A_\mu\right)\Phi^i\right|^2+m^2\left|\Phi^i\right|^2\,,
\label{2.4}
\end{equation}
where $F_{\mu\nu}=\partial_\mu A_\nu-\partial_\nu A_\mu$ and $m$ is the
mass of the complex scalar fields. 

If we rewrite
\begin{equation}
\phi(x)\equiv\frac{e}{\sqrt{N}}L\langle A_y\rangle\,,
\label{2.5}
\end{equation}
the classical action for $\phi$ becomes
\begin{equation}
\Gamma^{(0)}=N\int d^3x\frac{1}{2}\frac{1}{e^2L^2}(\partial_x\phi)^2\,.
\label{2.6}
\end{equation}

Obviously, the one-loop effective action is expressed with an overall
factor $N$%
\footnote{Apparently, the one-loop effect of quantization of the $U(1)$
field does not cause the effective action for $\phi$.}. Since the two-
and higher-loop correction is at most order $O(N^0)$ only the classical
and the one-loop action are effective in the limit $N\rightarrow\infty$.

Because of the coupling dependence in (\ref{2.6}) and the one-loop
potential which is proportional to $L^{-3}$, the typical scale for the
`thickness' of the kink turns out to be $L/(e^2L)^{1/2}$. Therefore
we may omit higher-derivative contribution to the kink if the
dimensionless quantity $e^2L$ is sufficiently small. Here we consider
only second-order derivative correction at one loop under the
assumption $e^2L\ll 1$.

To avoid the repetition of $N$, we use the action `per one charged
field' from now on.

The effective potential for r from one-loop contribution of complex scalar field in
massless limit can be found as\footnote{In this paper we consider the
fields satisfying the periodic boundary condition on $S^1$, $\Phi(y+L)
=\Phi(y)$.}
\begin{equation}
V(\phi)=\Gamma^{(1)}_0\large/ \int d^3x=\frac{2}{\pi L^3}
\sum_{l=1}^\infty\frac{\sin^2(l\phi/2)}{l^3}\,,
\label{2.7}
\end{equation}
where we set the energy of the vacuum at $\phi=0$ (mod $2\pi$) to zero,
since the vacuum energy independent of $\phi$ is irrelevant for our
present purpose.

The second-derivative term induced by the massless scalar field is written as
\begin{equation}
\Gamma^{(1)}_2=\int d^3x\frac{1}{48\pi}{\rm cosec}^2\frac{\phi}{2}\,
\frac{(\partial_x\phi)^2}{L}\equiv\int d^3x\frac{1}{2}
\frac{\gamma(\phi)}{e^2L^2}(\partial_x\phi)\,.
\label{2.8} 
\end{equation}

The quantum effect from fermions is also calculated by a similar method. The
effective potential induced by a single, massless Dirac fermion field is
\begin{equation}
V_f(\phi)=\frac{1}{\pi L^3}\sum_{l=1}^\infty\frac{\cos
l\phi-(-1)^l}{l^3}\,,
\label{2.9}
\end{equation}
and the second-derivative correction from a single, massless Dirac field is given
by
\begin{equation}
\Gamma^{(1)}_{2,f}=\int d^3x\frac{1}{96\pi}{\rm cosec}^2\frac{\phi}{2}\,
\frac{(\partial_x\phi)^2}{L}\,.
\label{2.10}
\end{equation}

In the appendix, the general expressions for the one-loop effects of scalars and
fermions with arbitrary masses at arbitrary temperature are found.

In the next section, we derive the kink solution in the presence of the derivative
correction.

\section{Quantum kinks}
In this section we explicitly solve the one-loop modified action to
obtain the kink solution. Now a short comment is in order. In pure
$U(1)$ gauge + {\it massless} fermion system, the vacuum is located at
$\phi=\pi$ (mod $2\pi$). Thus the topological soliton which
interpolates two vacua must pass through $\phi=0$ (mod $2\pi$), where
the second-derivative correction diverges. This exhibits the failure of
the expansion of effective action in terms of the number of derivatives
\cite{5}. It is also said that the Green's function in the momentum
space has singularities or non-analytic points \cite{6}.

For a gauge system with complex scalar fields, the difficulty does not appear
because the gradient of the field is rapidly reduced near $\phi=0$ for
the `tail' of the kink\footnote{Of course the derivative expansion
may be permitted for certain scalar-fermion mixed matter systems.}. We
will evaluate the contribution of the higher order in the derivative
expansion.

From now on, we consider the effective action from a scalar one loop.
The effective action we now consider is
\begin{equation}
\Gamma=L\int_{-\infty}^\infty
dx\left[\frac{1}{2}\frac{1+\gamma(\phi)}{e^2L^2}
(\partial_x\phi)^2+V(\phi)\right]\,,
\label{3.1}
\end{equation}
where $\gamma$ and $V$ are given by (\ref{2.8}) and (\ref{2.7}).

Here we want static solutions, and then we suppressed the integration by time
variable, while we integrated out in terms of the coordinate $y$, in
(\ref{3.1}). 

We can simplify the action by using the coordinate $z$ defined as
\begin{equation}
z\equiv\sqrt{e^2L}\frac{x}{L}\,,
\label{3.2}
\end{equation}
and we get
\begin{equation}
\Gamma=\frac{1}{\sqrt{e^2L}L}\tilde{\Gamma}=\frac{1}{\sqrt{e^2L}L}
\int_{-\infty}^\infty dz \left[\frac{1}{2}(1+\gamma(\phi))
(\partial_x\phi)^2+\tilde{V}(\phi)\right]\,,
\label{3.3}
\end{equation}
where $\tilde{V}(\phi)=L^3V(\phi)$.

\begin{figure}[ht]
\begin{center}
\includegraphics[width=6cm]{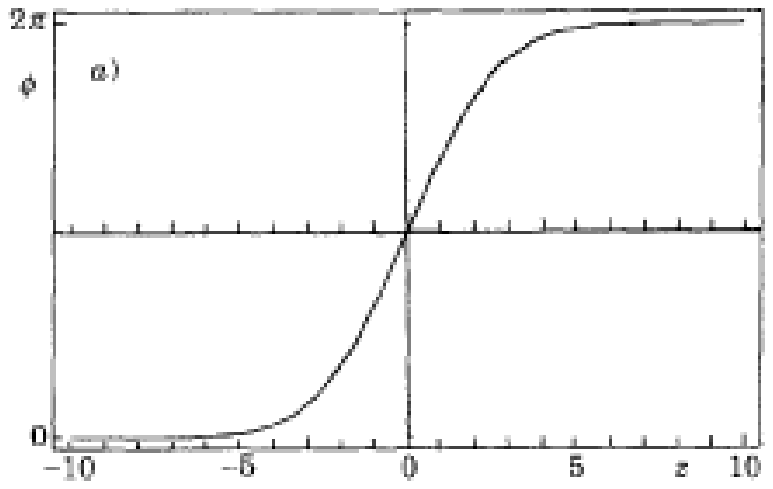}
\includegraphics[width=6cm]{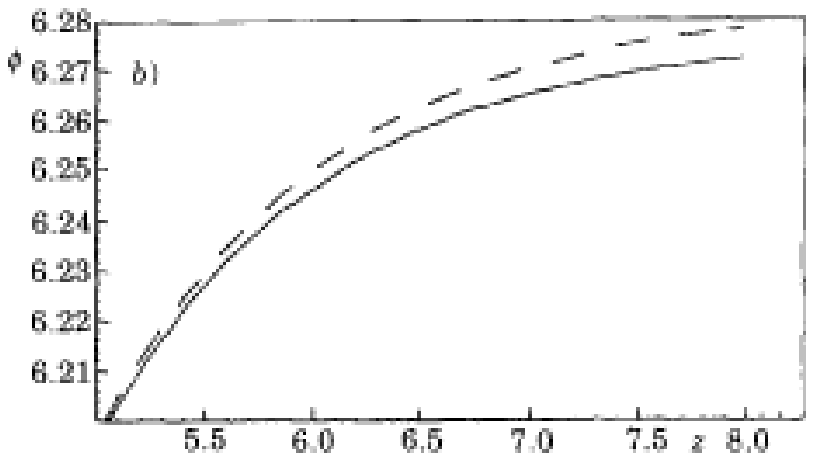}
\caption{a) A view of a quantum kink. b) The solid line shows the kink
in the system governed by the effective action (\ref{3.1}). The dashed
line shows the kink in the same system but where the derivative
correction $\gamma$ is omitted.}
\label{f1}\end{center}
\end{figure}

The kink solution with the boundary conditions $\phi(-\infty)=0$ and
$\phi(\infty)=2\pi$ can be easily obtained from this action. The shape
of kink is shown in fig. 1a) for $e^2L = 0.01$.
To see the effect of derivative correction, we also show the kink for the system
governed by the action (\ref{3.1}) but without $y$. The difference in
the profile is very small and thus a small portion of fig.~1a) is shown
in fig.~1b) with an appropriate magnification. The solid line shows the
piece of the kink in the system governed by the effective action
(\ref{3.1}), while the dashed line shows the kink in the same system but
with the derivative correction $\gamma$ omitted.

The tail of the kink approaches the vacuum value as $z^{-1}$ in the
presence of the second-derivative correction. Without the derivative
correction, $\phi$ is damped exponentially near the vacua. The
derivative correction makes the kink less steeper;
nevertheless no
singular behaviour is found, and thus the higher order in the
derivative expansion is expected to be suppressed by the power of $e^2
L$. 

\begin{figure}[ht]
\begin{center}
\includegraphics[width=6cm]{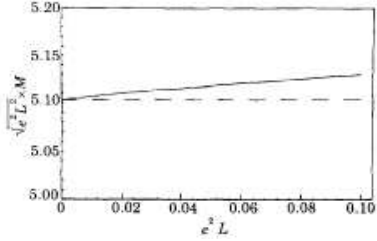}
\caption{The kink mass in the cases with and without the derivative
correction, as a function of $e^2L$. The solid line and dashed line
correspond to the kink mass with and without the second-derivative
correction. Note that the mass is scaled by $1/\sqrt{e^2L^3}$. 
}
\label{f2}\end{center}
\end{figure}

The mass of the kink $M$ is given by $\Gamma(\phi_{kink})$. The
difference between the kink mass with and without the derivative
correction as a function of $e^2L$ is shown in fig.~2. The derivative
correction reduces the kink mass.

The mass of the kink is expressed approximately as
\begin{equation}
M=\frac{1}{\sqrt{e^2L^3}}(5.104+0.26 e^2L)\,.
\label{3.4} 
\end{equation}

Note that this expression is derived for one massless complex scalar
field. For $N$ scalar fields, the mass is the above result multiplied by
$N$. It is a pleasure to point out that the leading term for the kink
mass takes a very similar form as the monopole case, where the mass is
given by the typical mass scale in the system divided by the absolute
value of the gauge coupling.

\section{Quantum kinks at finite temperature}
Now, we consider the finite-temperature effect. First, we derive the
effective action for r in the high-temperature limit.
The effective potential which comes from the one-loop effect of a complex scalar
field with mass m at high temperature $T$ ($\gg L^{-1}$) turns out to be
\begin{equation}
V_{high~T}(\phi)=\frac{4T}{\pi}\sum_{l=1}^\infty\frac{m}{Ll}
K_2(mLl)\sin^2\frac{l\phi}{2}\,,
\label{4.1}
\end{equation}
where $K_2$ is the modified Bessel function of the second kind and the
term independent of $\phi$ is dropped. (See also the appendix.)

The correction to the kinetic term for r at high temperature is given by
\begin{equation}
\Gamma^{(1)}_{2, high~T}\frac{T}{8}\int d^3x\sum_{l=-\infty}^\infty
\frac{(2\pi l+\phi)^2}{((2\pi l+\phi)^2+m^2)^{5/2}}(\partial_x\phi)^2\,.
\label{4.2}
\end{equation}
(See also the appendix.) The typical scale of the gradient of the kink configuration
becomes $(e^2 TL^2)^{1/2}L$ at high temperature and small $m$, instead
of $(e^2L)^{1/2}L$ at zero temperature.

The derivative correction makes the tail of kink longer in general.
$\phi$ approaches the vacuum value as $x^{-1}$ in the limit of the
infinitesimal mass for the scalar field, as the zero temperature case.

However, even if the kink solution exists, the mass of the kink diverges in the
massless limit for the scalar field. This is due to the integration in the region of the
long tail of the kink. The mass of the kink inflates for the case with the scalar field
with small mass at high temperature, as $(T/m)^{1/2}$. The divergence
may originate from the infrared behaviour of the field in the
low-dimensional space.

\section{Non-Abelian case}
In this section, we consider the one-loop quantum effect of $SU(2)$
Yang-Mills field. The classical Lagrangian is given by
\begin{equation}
L_{YM}=\frac{1}{4}\sum_{a=1}^3(F^a_{\mu\nu})^2\,,
\label{5.1} 
\end{equation}
where the gauge field strength is defined as
\begin{equation}
F^a_{\mu\nu}=\partial_\mu A^a_\nu-\partial_\nu
A^a_\mu-e\varepsilon^{abc} A_\mu^b A_\nu^c\,.
\label{5.2}
\end{equation}

Then the kink consists of a component of the gauge field, say,
$\langle A^3_y\rangle$.

We take the gauge-fixing term as
\begin{equation}
L_{g.f.}=\frac{1}{2\alpha}\sum_{a=1}^3(B^a)^2\,,
\label{5.3}
\end{equation}
with
\begin{eqnarray}
B^1&=&\partial_iA_i^1+\alpha(\partial_y A_y^1+\langle A_y^2\rangle
A_y^2)+2U^i A_i^1\,,
\label{5.4a}\\
B^2&=&\partial_iA_i^2+\alpha(\partial_y A_y^2-\langle A_y^2\rangle
A_y^1)+2U^i A_i^2\,,
\label{5.4b}\\
B^3&=&\partial_iA_i^3+\alpha(\partial_y A_y^3)+2U^i A_i^3\,,
\label{5.4c}
\end{eqnarray}
where $A_y^3$ denotes the deviation from the background field, $i=\{t,
x\}$ and $U_x=\partial_x\langle A_y^3\rangle(\langle
A_y^3\rangle)^{-1}$. $\alpha$ is the gauge parameter we set to unity
hereafter. 

Combining (\ref{5.1}) and (\ref{5.3}), we find that the one-loop
effective action is written in the form
\begin{equation}
\Gamma^{(1)}=\frac{1}{2}{\rm tr~} \ln\Delta_{YM}-{\rm tr~}
\ln\Delta_{gh}\,,
\label{5.5}
\end{equation}
where
\begin{eqnarray}
\Delta_{YM}^{\mu\nu}&=&-\delta^{\mu\nu}(D_\lambda)^2+2(U^\mu
\partial^\nu-U^\nu\partial^\mu)-[\partial^\mu U^\nu+\partial^\nu U^\mu
]+U^\mu U^\nu\,,
\label{5.6)}\\
\Delta_{gh}&=&-(D_\lambda)^2-2U^\mu\partial_\mu\,.
\label{5.7}
\end{eqnarray}

Here the $SU(2)$ indices are suppressed and the covariant derivative is
defined as
\begin{eqnarray}
D_i&=&\delta^{ab}\partial_i\,,
\label{5.8}\\
D_y&=&\delta^{ab}\partial_y+e\varepsilon^{abc}\langle A_y^c\rangle\,.
\label{5.9}
\end{eqnarray}
$U_y\equiv 0$ is added as a dummy.

After the calculation is done similarly to ref. \cite{4}, we find the
second-order derivative correction to the effective action for
$\phi\equiv eL\langle A_y^3\rangle$ at one-loop order at zero
temperature:
\begin{equation}
\Gamma^{(1)}_2=-\int d^3x \frac{11}{48\pi}{\rm cosec}^2\frac{\phi}{2}
\frac{(\partial_x\phi)^2}{L}\,.
\label{5.10} 
\end{equation}
The sign of the above correction is the opposite sign of the classical term. Thus one
may suspect the instability of the vacuum. The inclusion of the other matter field may
ensure the stability. In addition, the two-loop contribution may be important (since
one cannot argue for omission of the higher loops by large $N$, unlike
the previous case for scalar fields), so we cannot declare a definite
statement.

\section{Conclusion and remarks}
In the present paper we have studied quantum kinks in gauge theories on
$R^2\otimes S^1$. Since the kink solutions should be treated as the
spatially dependent background, we have calculated the effective action
by the method introduced by Moss, Toms and Wright \cite{4}. For scalar
QED, the effective action has been calculated and the kink solution has
been obtained numerically.

At high temperature, the kink mass diverges in the limit of the infinitesimal mass
for the complex scalar field, though the kink solution exists. This fact may indicate
the infrared problem in such a low-dimensional system.

For the pure $SU(2)$ Yang-Mills system, the possible instability of the
gauge vacuum due to the derivative correction has been found. The
instability can be cured by adding matter fields. Moreover, it is
possible that the higher-loop effect may change the effective action
largely.

We have presented an example in which the derivative correction has a sense in
the calculable quantity (=kink mass). We have not yet known other
examples for applications of the derivative correction to concrete
calculation of physical quantities, though the importance of the
derivative correction has been advocated in the literature on the
cosmological phase transitions. It may be necessary to investigate the
evaluation of the derivative correction by other methods than the
expansion in terms of the number of the derivatives. We also study the
singularity of the Green's function in the momentum space in general.

Finally, we hope that the kink we have studied has some relevance to the
low-dimensional physics.

\section*{APPENDIX A}
In this appendix we exhibit the effective potential $V(\phi)$ and the
second-derivative correction $\gamma(\phi)$ for the scalar and Dirac
fields for the vacuum $U(1)$ gauge field as well as for the $SU(2)$
Yang-Mills field for the vacuum gauge field. 

\subsection*{A) Complex scalar fields.}

For a complex scalar field with mass $m$ at temperature $T$, we find
\begin{equation}
V(\phi)=\frac{\sqrt{2}m^3}{\pi^{3/2}}\left(\sum_{n=-\infty}^{\infty}
\sum_{l=-\infty}^\infty\right)'
\frac{K_{3/2}(z_{nl})}{z_{nl}^{3/2}}\sin^2\frac{l\phi}{2}\,,
\label{A.1}
\end{equation} 
where $K_{3/2}(z)$ is the modified Bessel function of the
second
kind, $z_{nl}\equiv m\sqrt{(n/T)^2+(Ll)^2}$ and the prime
on sums indicates that the term with $n =l=0$ is to be omitted. We
also omit the vacuum energy which is independent of $\phi$.

And the
second-derivative correction is found to be
\begin{equation}
\gamma(\phi)=\frac{e^2TL^2}{4}\sum_{n=-\infty}^\infty
\sum_{l=-\infty}^\infty
\frac{(2\pi l+\phi)^2}{
[(2\pi TLn)^2+(2\pi l+\phi)^2+(mL)^2]^{5/2}}\,.
\label{A.2}
\end{equation}

The results for massless fields at zero and high-temperature in the
text are obtained by taking the limit $m\rightarrow 0$ and the
corresponding limit on $T$. 

\subsection*{B) Dirac fermion fields.}
For a Dirac fermion field with mass $m$ at temperature $T$, we find
\begin{equation}
V_f(\phi)=\frac{m^3}{\sqrt{2}\pi^{3/2}}\left(\sum_{n=-\infty}^\infty
\sum_{l=-\infty}^\infty\right)'(-1)^n\frac{K_{3/2}(z_{nl})}{z_{nl}^{3/2}}
(\cos l\phi-(-1)^l)\,,
\label{A.3}
\end{equation}
where $K_{3/2}(z)$ is the modified Bessel function of the second kind,
$z_{nl}\equiv m\sqrt{(n/T)^2+(Ll)^2}$ and the prime on sums indicates
that the term with $n=l=0$ is to be omitted. We also omit the vacuum
energy which is independent of $\phi$.
 
And the second-derivative
correction is found to be
\begin{equation}
\gamma_f(\phi)=\frac{e^2TL^2}{4}\sum_{n=-\infty}^\infty
\sum_{l=-\infty}^\infty
\frac{(2\pi TL(n+1/2))^2+(mL)^2}{
[(2\pi TL(n+1/2))^2+(2\pi l+\phi+(mL)^2]^{5/2}}\,.
\label{A.4}
\end{equation}

The results for massless fields at zero temperature in the text are obtained by
taking the limit $m, T\rightarrow 0$.

\subsection*{C) $SU(2)$ Yang-Mills fields (at zero temperature).}
For $SU(2)$ Yang-Mills fields at zero temperature, we find
\begin{equation}
V_{YM}(\phi)=\frac{2}{\pi}\sum_{l=1}^\infty
\frac{1}{(Ll)^3}\sin^2\frac{l\phi}{2}\,,
\label{A.5}
\end{equation}
where we omit the vacuum energy which is independent of $\phi$.

And the second-derivative correction at zero temperature is found to be
\begin{equation}
\gamma_{YM}(\phi)=-\frac{11e^2L}{24\pi}{\rm cosec}^2\frac{\phi}{2}\,.
\label{A.6}
\end{equation}



\begin{thebibliography}{99}
\bibitem{1} Y. Hosotani, Phys. Lett. {\bf B126} (1983) 309;
D. J. Toms, Phys. Lett. {\bf B126} (1983) 445;
E. Witten, Nucl. Phys. {\bf B258} (1985) 75. 

\bibitem{2} K. Lee, R. Holman and E. W. Kolb, Phys. Rev. Lett. {\bf 59}
(1987) 1069;
B.-H. Lee et al., Phys. Rev. Lett. {\bf 60} (1988) 2231;
M. Tomiya, J. Phys. {\bf G14} (1988) L153. 

\bibitem{3} A. Higuchi and L. Parker, Mod. Phys. Lett. {\bf A5} (1990)
2251.

\bibitem{4} I. Moss, D. Toms and A. Wright, Phys. Rev. {\bf D46}
(1992) 1671. 

\bibitem{5} S. G. Naculich, Phys. Rev. {\bf D46} (1992) 5487;
S. Dodelson and B.-A. Gradwohl, Nucl. Phys. {\bf B400} (1993) 435. 

\bibitem{6} P. Elmfors, K. Enqvist and I. Vilja, 
Nucl. Phys. {\bf B422} (1994) 521.

\end{thebibliography}
\end{document}